\documentclass[aip,preprint]{revtex4-1}
\usepackage{graphicx}
\usepackage[section]{placeins}
 \usepackage{float}

\begin{document}


\title{On the thermodynamic properties of fictitious identical particles and the application to fermion sign problem}

\author{Yunuo Xiong}
\affiliation{Center for Fundamental Physics and School of Mathematics and Physics, Hubei Polytechnic University, Huangshi 435003, People’s Republic of China}

\affiliation{College of Science, Zhejiang University of Technology, Hangzhou 31023, People’s Republic of China}

\author{Hongwei Xiong}
\email{xionghw@zjut.edu.cn}

\affiliation{Center for Fundamental Physics and School of Mathematics and Physics, Hubei Polytechnic University, Huangshi 435003, People’s Republic of China}

\affiliation{College of Science, Zhejiang University of Technology, Hangzhou 31023, People’s Republic of China}


\begin{abstract}
By generalizing the recently developed path integral molecular dynamics for identical bosons and fermions, we consider the finite-temperature thermodynamic properties of fictitious identical particles with a real parameter $\xi$ interpolating continuously between bosons ($\xi=1$) and fermions ($\xi=-1$). Through general analysis and numerical experiments we find that the average energy may have good analytical property as a function of this real parameter $\xi$, which provides the chance to calculate the thermodynamical properties of identical fermions by an extrapolation with a simple polynomial function after accurately calculating the thermodynamic properties of the fictitious particles for $\xi\geq 0$. Using several examples, it is shown that our method can efficiently give accurate energy values for 
finite-temperature fermionic systems. Our work provides a chance to circumvent the fermion sign problem for some quantum systems.
\end{abstract}


\maketitle

\section{Introduction}

The partition function is the starting point of quantum statistical physics. Based on Feynman's idea of path integral \cite{feynman,kleinert,Tuckerman}, we can transform the partition function of many-particle systems into classical interacting ring polymers \cite{chandler,Parrinello,Miura,Cao,Cao2,Jang2,Ram,Kinugawa,Roy,Roy1,Roy2,Roy3,Kinugawa1,Kinugawa2,Roy4,Roy5,Roy6,Poly,Craig,Braa,Haber,Thomas} and use Monte Carlo \cite{CeperRMP,boninsegni1,boninsegni2,Dornheim,DornheimMod} or molecular dynamics (MD) \cite{Hirshberg,HirshbergFermi,Deuterium,Xiong,Xiong2,Xiong3,Xiong4,Xiong5} approaches to sample the probability distribution in the partition function in order to obtain the thermodynamic properties such as the average energy. Usually, the exchange symmetry of identical bosons or the exchange antisymmetry of identical fermions will lead to extra difficulty in transforming the partition function into classical interacting ring polymers. Fortunately, a recursion formula \cite{Hirshberg,HirshbergFermi} is found to transform the partition function into classical interacting ring polymers for both identical bosons and fermions, so that the path integral molecular dynamics (PIMD) can be used to study the thermodynamic properties of identical particles. For identical bosons, PIMD has been successfully applied to predict a supersolid phase in high-pressure deuterium \cite{Deuterium}. However, for identical fermions, due to the exchange antisymmetry of fermions, the distribution function in the partition function is not positive definite, rendering existing sampling methods ineffective. Despite the development of various approaches to alleviate the fermion sign problem \cite{DornheimMod,HirshbergFermi}, there are still insurmountable difficulties for the general solution of fermion sign problem \cite{Alex,ceperley,troyer,loh,lyubartsev,vozn,Science,Wu,Umrigar,Li,Wei,Yao1,Yao2}. 

In this work, we consider the finite-temperature thermodynamic properties of fictitious identical particles with a real parameter $\xi$ interpolating continuously between bosons ($\xi=1$) and fermions ($\xi=-1$). The property of the fictitious identical particle is chosen such that the recursion formula for the partition function to consider identical bosons or fermions in PIMD can be generalized to this fictitious identical particle. 

We will focus on the calculation of the average energy $E(\xi)$ in this work. There is a sign problem for $\xi<0$ which means that the traditional method to calculate $E(\xi)$ for $\xi<0$ has a computational scaling of $O(e^N)$ with $N$ the number of fermions, while we can calculate accurately and efficiently the thermodynamic properties of the fictitious identical particles for different parameters $\xi$ satisfying $\xi\geq 0$ with a computational scaling of $O(N^3)$. 
We find by general analysis and numerical simulation that $E(\xi)$ may have good analytical property for $\xi\geq -1$. This provides the chance to get the average energy $E(\xi=-1)$ for fermions by an extrapolation of analytical continuation, after we calculate the average energy $E(\xi)$ for different positive $\xi$ without suffering from sign problem. We used numerical experiments to verify our idea for different fermionic systems, and find that our result of the energy for fermions agrees with that of the traditional method needing much more  computational resources.  The method of the present work to circumvent fermion sign problem may have potential applications for many of the problems that encounter difficulties due to fermion sign problem.

\section{Partition function of fictitious identical particles and fermion sign problem}

\subsection{Partition function of fictitious identical particles}

We consider the following Hamiltonian for $N$ fictitious identical particles
\begin{equation}
\hat H=\frac{1}{2m}\sum_{l=1}^N\hat \textbf p_l^2+\hat V(\textbf r_1,\cdots,\textbf r_N).
\end{equation}
Here $V(\textbf r_1,\cdots,\textbf r_N)$ includes both the external potential and inter-particle interactions. 
The partition function of the fictitious identical particles is
\begin{equation}
Z(\beta,\xi)=Tr(e^{-\beta \hat H}).
\end{equation}
Here $\xi$ in $Z(\beta,\xi)$ is a parameter for the fictitious identical particles, which will be introduced in due course. The average energy is
\begin{equation}
E(\beta,\xi)=-\frac{\partial\ln Z(\beta,\xi)}{\partial \beta}.
\end{equation}

In coordinate representation to consider the trace in the above partition function, we have
\begin{equation}
Z(\beta,\xi)\sim\sum_{p\in S_N}\xi^{N_p}\int d\textbf{r}_1d\textbf{r}_2\cdots d\textbf{r}_N\left<p\{\textbf{r}\}|e^{-\beta \hat H}|\{\textbf{r}\}\right>.
\label{Xipartition}
\end{equation}
Here $\beta=1/k_B T$, with $k_B$ being the Boltzmann constant and $T$ being the system temperature. $\{\textbf{r}\}$ denotes $\{\textbf{r}_1,\cdots,\textbf{r}_N\}$.  $S_N$ represents the set of $N!$ permutation operations denoted by $p$. The factor $\xi^{N_p}$ is due to the exchange effect of identical particles, with $N_p$ a number defined to be the minimum number of times for which pairs of indices must be interchanged in permutation $p$ to recover the original order. The above expression of the partition function gives a unified description of identical bosons and fermions, because 
$\xi=+1$ for boson partition function, while $\xi=-1$  for fermion partition function. We note that, from a purely mathematical point of view,  $\xi$ can be any complex number in the above partition function, and the situation of $\xi\neq \pm 1$ denotes a fictitious identical particle. In this work, we consider the situation of real $\xi$ to circumvent the fermion sign problem.

\subsection{Fermion sign problem and traditional method}

Using $e^{-\beta\hat H}=e^{-\Delta\beta\hat H}\cdots e^{-\Delta\beta\hat H}$ with $\Delta\beta=\beta/P$ and the technique of path integral, the partition function $Z(\beta,\xi=\pm 1)$ can be mapped as a classical system of interacting ring polymers. 
In recent works by Hirshberg et al. \cite{Hirshberg,HirshbergFermi}, a recursion formula is found to calculate the partition function for both bosons ($\xi=1$) and fermions ($\xi=-1$). Based on the path integral ring polymer, the fermion partition function ($Z_F$) and boson partition function ($Z_B$) are expressed as an integral for an analytical function of the coordinates of interacting ring polymers 
\begin{equation}
Z_{F/B}\sim\int d\textbf{R}_1\cdots d\textbf{R}_N e^{-\beta U_{F/B}^{(N)}}.
\end{equation}
Here the integrand is a function of a set of ring polymer coordinates $(\textbf{R}_1,\cdots,\textbf{R}_N)$, with $\textbf{R}_i=(\textbf{r}_i^1,\cdots,\textbf{r}_i^P)$ corresponding to $P$ ring polymer coordinates for the $i$th particle. 
A recursion formula is found to give the expression of $U_F^{(N)}$ for fermions and $U_B^{(N)}$ for bosons. For bosons, $U_B^{(N)}$ is always real and $e^{-\beta U_{B}^{(N)}}$ may be sampled for large boson systems \cite{Hirshberg,Deuterium,Xiong,Xiong2}. However, because of the factor $(-1)^{N_p}$ for fermions, $U_F^{(N)}$ is a complex function, which can not be sampled with available methods. 

To deal with the situation of complex function $U_F^{(N)}$, in the usual traditional method \cite{DornheimMod,HirshbergFermi} to deal with the fermion sign problem, the fermion partition function may be rewritten as \cite{HirshbergFermi}
\begin{equation}
Z_{F}\sim\int d\textbf{R}_1\cdots d\textbf{R}_N e^{-\beta U_{B}^{(N)}}e^{\beta(U_B^{(N)}-U_F^{(N)})}.
\end{equation}
In this case, we may still use molecular dynamics to sample $e^{-\beta U_{B}^{(N)}}$, while the energy is $E(\beta)=\left<\epsilon s\right>_B/{\left<s\right>_B}$.
Here $s=e^{\beta(U_B^{(N)}-U_F^{(N)})}$ and $\epsilon$ is the estimator for energy. The average $\left<\cdots\right>_B$ is about the samples based on the partition function $Z_B$ for bosons.

However, because of the factor $s$, it is hard to calculate the energy, both in path integral Monte Carlo \cite{ceperley,Alex,Science,CeperRMP,boninsegni1,boninsegni2,Dornheim,DornheimMod} and path integral molecular dynamics \cite{HirshbergFermi,Xiong2}. General consideration shows that the simulation becomes exponentially hard to converge with increasing numbers of fermions and decreasing temperatures \cite{ceperley}. This is known as the infamous fermion sign problem \cite{ceperley,Alex,loh,troyer,lyubartsev,vozn,Science,Wu,Umrigar,Li,Wei,Yao1,Yao2}.

\section{Recursion formula and energy estimator for fictitious identical particles}

From the expression of $Z(\beta,\xi)$, it is mathematically legitimate that we extend two discrete values $\xi=\{-1,1\}$ in $Z(\beta,\xi)$ to the general case that $\xi$ becomes a continuous real variable. To calculate the energy $E(\beta,\xi)$ for fictitious identical particles, we turn to consider the partition function $Z(\beta,\xi)$ given by Eq. (\ref{Xipartition}). 

\subsection{Recursion formula for $Z(\beta,\xi)$}

We consider $N$ fictitious identical particles, so that the normalized position eigenstates are
\begin{equation}
\left|N_\xi\right>=\frac{A(\xi)}{\sqrt{N!}}\sum_{p\in S_N}\xi^{N_p}\left|p\{\textbf{r}\}\right>.
\end{equation}
$S_N$ represents the set of $N!$ permutation operations denoted by $p$. $A(\xi)$ is a normalization constant.

We have the unit operator
\begin{equation}
\hat I_\xi=\frac{A^2(\xi)}{N!}\int d\textbf{r}_1d\textbf{r}_2\cdots d\textbf{r}_N \left|N_\xi\right>\left<N_\xi\right|.
\end{equation}
We also have another unit operator
\begin{equation}
\hat I=\int d\textbf{r}_1d\textbf{r}_2\cdots d\textbf{r}_N \left|\{\textbf{r}\}\right>\left<\{\textbf{r}\}\right|.
\end{equation}

By dividing $\beta$ into $P$ segments, with $\Delta\beta=\beta/P$, and using the identity
\begin{equation}
e^{-\beta \hat H}\hat I_\xi=e^{-\Delta\beta \hat H}\hat I e^{-\Delta\beta \hat H}\hat I\cdots \hat I e^{-\Delta\beta \hat H}\hat I_\xi,
\end{equation}
we can obtain the discrete form for the partition function by inserting the definition for the identity operators from the above equations. After that is done, the partition function is a function of a set of ring polymer coordinates $(\textbf{R}_1,\cdots,\textbf{R}_N)$, with $\textbf{R}_i=(\textbf{r}_i^1,\cdots,\textbf{r}_i^P)$ corresponding to $P$ ring polymer coordinates for the $i$th particle. Explicitly, after considering all the permutation terms in Eq. (\ref{Xipartition}),  generalizing the technique of the recursion formula discovered by Hirshberg et al. \cite{Hirshberg,HirshbergFermi}, the partition function can be written as
\begin{equation}
Z(\beta,\xi)\sim\int d\textbf{R}_1\cdots d\textbf{R}_N e^{-\beta U_\xi^{(N)}}.
\label{Zxi}
\end{equation}
$U_\xi^{(N)}$ is given by
\begin{equation}
U_\xi^{(N)}=-\frac{1}{\beta}\ln W_\xi^{(N)}+\frac{1}{P}\sum_{j=1}^P V\left(\textbf{r}_1^j,\cdots,\textbf{r}_N^j\right),
\label{UFN}
\end{equation}
and $W_\xi^{(N)}$ is
\begin{equation}
W_\xi^{(N)}=\frac{1}{N}\sum_{k=1}^N\xi^{k-1}e^{-\beta E_N^{(k)}}W_\xi^{(N-k)}.
\label{WFN}
\end{equation}
In addition,
\begin{equation}
E_N^{(k)}=\frac{1}{2}m\omega_P^2\sum_{l=N-k+1}^N\sum_{j=1}^P\left(\textbf r_l^{j+1}-\textbf{r}_l^j\right)^2.
\end{equation}
Here $\textbf r_l^{P+1}=\textbf r_{l+1}^1$, except for $l=N$ for which $\textbf r_N^{P+1}=\textbf r_{N-k+1}^1$.
In addition, $\omega_P=\sqrt{P}/\beta\hbar$.

In the limit of $\xi\rightarrow 0^+$, in the partition function given by (\ref{Xipartition}), only one term with $N_p=0$ should be reserved. This is the partition function for distinguishable particles. Hence, in this work, the numerical calculation for $\xi=0$ should be given special consideration, which refers specifically to distinguishable particles.

\subsection{Energy estimator and molecular dynamics}

In order to perform molecular dynamics, we need the potential function and its gradient. From the fact that $W_\xi^{(N)}=e^{-\beta V_\xi^{(N)}}$, we can evaluate the potential $V_\xi^{(N)}$ recursively as
\begin{equation}
V_\xi^{(\alpha)}=-\frac{1}{\beta}\log{\frac{1}{\alpha}\sum_{k=1}^\alpha\xi^{k-1}e^{-\beta (E_\alpha^{(k)}+V_\xi^{(\alpha-k)})}}.
\end{equation}
Taking gradient, we see that the forces are given by
\begin{equation}
-\nabla_{\mathbf r_l^j}V_\xi^{(\alpha)}=-\frac{\sum_{k=1}^\alpha\xi^{k-1}[\nabla_{\mathbf r_l^j}E_\alpha^{(k)}+\nabla_{\mathbf r_l^j}V_\xi^{(\alpha-k)}]e^{-\beta(E_\alpha^{(k)}+V_\xi^{(\alpha-k)})}}{\sum_{k=1}^\alpha\xi^{k-1} e^{-\beta(E_\alpha^{(k)}+V_\xi^{(\alpha-k)})}}.
\end{equation}

From the relation between $E(\beta,\xi)$ and $Z(\beta, \xi)$, we have
\begin{equation}
E(\beta,\xi)=\left<\epsilon(\beta,\xi)\right>,
\end{equation}
with the energy estimator $\epsilon(\beta,\xi)$ given by
\begin{equation}
\epsilon(\beta,\xi)=\frac{PdN}{2\beta}+\frac{U_\xi^{(N)}}{P}+V_\xi^{(N)}+\beta\frac{\partial V_\xi^{(N)}}{\partial\beta}.
\end{equation}
Here $d$ is the spatial dimensions.
$V_\xi^{(N)}+\beta\frac{\partial V_\xi^{(N)}}{\partial\beta}$ may be evaluated as
\begin{equation}
V_\xi^{(N)}+\beta\frac{\partial V_\xi^{(N)}}{\partial\beta}=\frac{\sum_{k=1}^N\xi^{k-1}[V_\xi^{(N-k)}+\beta\frac{\partial V_\xi^{(N-k)}}{\partial\beta}-E_N^{(k)}]e^{-\beta(E_N^{(k)}+V_\xi^{(N-k)})}}{\sum_{k=1}^N\xi^{k-1}e^{-\beta(E_N^{(k)}+V_\xi^{(N-k)})}},
\end{equation}
with $V_\xi^{(0)}+\beta\frac{\partial V_\xi^{(0)}}{\partial\beta}=0$.
\par
We should take care to ensure numerical stability when evaluating the above formulas. For the potential, except for the case of $\xi=0$, a numerically stable way to calculate potential is the following
\begin{equation}
V_\xi^{(\alpha)}=\frac{1}{\beta}(\tilde E-\log{\sum_{k=1}^\alpha e^{(k-1)\log{\xi}-\beta (E_\alpha^{(k)}+V_\xi^{(\alpha-k)})+\tilde E}}+\log{\alpha}),
\end{equation}
where $\tilde E$ is chosen as
\begin{equation}
\tilde E=-\max_k [(k-1)\log{\xi}-\beta (E_\alpha^{(k)}+V_\xi^{(\alpha-k)})].
\end{equation}

\section{General behavior of fictitious identical particles} 
\label{fullbehavior}

We expect that $E(\beta,\xi)$ calculated from  $Z(\beta,\xi)$ is a monotonic analytic function about $\xi$ because of two reasons. 

(1) Because the operator $e^{-\beta\hat H}$ is hermitian, it is clear that $Z(\beta,\xi)$ is a real function of $\beta$ and $\xi$ even for $\xi=-1$. From the expression of the partition function (\ref{Xipartition}), the partition function is a polynomial function of $\xi$. Hence, we expect that the average energy is also an analytical function of $\xi$. 

(2) $\xi$ denotes an exchange interaction for identical particles. For $\xi=1$ (bosons), it gives an equivalently attractive interaction, while it gives an equivalently repulsive interaction \cite{Griffiths} for $\xi=-1$ (fermions). This means that $E(\beta,\xi)$ is a monotonic increasing function of $\xi$, as $\xi$ decreases from $1$ to $-1$, at least for the usual quantum system.

It is not the purpose of the present work to give a rigorous mathematical proof for the above properties. We will confirm these properties based on several examples by calculating $E(\beta,\xi)$ for different $\xi$ with the traditional method.

In the traditional method, the partition function is written as
\begin{equation}
Z(\beta,\xi)\sim\int d\textbf{R}_1\cdots d\textbf{R}_N e^{-\beta U_{\xi=1}^{(N)}}e^{\beta(U_{\xi=1}^{(N)}-U_{\xi}^{(N)})}.
\end{equation}
In this case, we may use molecular dynamics to sample $e^{-\beta U_{\xi=1}^{(N)}}$, while the average energy is 
\begin{equation}
E(\beta,\xi)=\frac{\left<\epsilon(\beta,\xi) s(\beta,\xi)\right>_{\xi=1}}{\left<s(\beta,\xi)\right>_{\xi=1}}.
\end{equation} 
Here $s(\beta,\xi)=e^{\beta(U_{\xi=1}^{(N)}-U_{\xi}^{(N)})}$ and $\epsilon(\beta,\xi)$ is the estimator for energy. The average $\left<\cdots\right>_{\xi=1}$ is about the samples based on the partition function for $\xi=1$ (bosons).

For $\xi\geq 0$, because $U_{\xi}^{(N)}$ is a real function, there is no sign problem to prevent accurate numerical simulation of the energy. However, for $\xi< 0$, $U_{\xi}^{(N)}$ is a complex function, which will lead to the sign problem. To calculate accurately the energy for $-1\leq \xi\leq 1$, we had to choose only a few fermions in the following examples, so that the sign problem is not severe. We consider fictitious identical particles in a two-dimensional harmonic trap $\frac{1}{2}m\omega^2(x^2+y^2)$ with the units of $\hbar=1,\omega=1,m=1$.

\begin{figure}[htbp]
\begin{center}
 \includegraphics[width=0.9\textwidth]{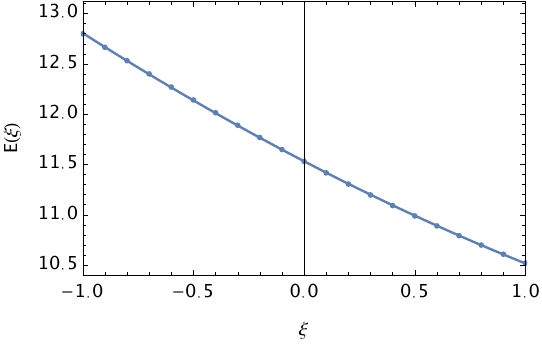} 
\caption{\label{4intFull} The circle is the energy by traditional method for Coulomb interaction with $N=4, \beta=1$, and $\lambda=0.5$. The solid line is the parabolic function for fitting. By adding more parameters with $a_3 \xi^3$ and $a_4 \xi^4$, we do not notice perceptible difference. Hence, we do not add those extra curves in this figure.}
\end{center}
\end{figure}

For Coulomb interaction, we have
\begin{equation} 
V_{int}=\sum_{l<j}^N \frac{\lambda}{{|{\textbf r}_l-{\textbf r}_j|}}.
\end{equation}
For $N=4, \beta=1$ and $\lambda=0.5$, in Fig. \ref{4intFull}, we give the numerical results for the whole region $-1\leq \xi\leq 1$ by using the traditional method to deal with the sign problem. In our simulations, 12 beads and 3 independent trajectories each consisting of $5 \times 10^6$ MD steps with separate Nóse-Hoover thermostat \cite{Nose1,Nose2,Hoover,Martyna,Jang,Schlessinger} are used to assure convergence.

In Fig. \ref{4intFull}, we give the following parabolic function 
\begin{equation}
E(\beta,\xi)=a_0+a_1\xi+a_2\xi^2
\end{equation}
to fit the data in the whole region, and agreement is found. This shows that $E(\beta,\xi)$ is a simple monotonic analytical function about $\xi$ in the whole region connecting bosons and fermions.

We consider the following mean deviation for the fitting defined by
\begin{equation}
\Delta=\sqrt{\frac{\sum_{j=1}^M(E_j-E(\beta,\xi_j))^2}{M}}.
\end{equation}
Here $M$ is the total number of data. $E_j$ is the energy of $\xi_j$ in the numerical calculation. For the example of Fig. \ref{4intFull}, $\Delta=0.0032$, which confirms the agreement of the fitting with parabolic function.

\begin{figure}[htbp]
\begin{center}
 \includegraphics[width=0.9\textwidth]{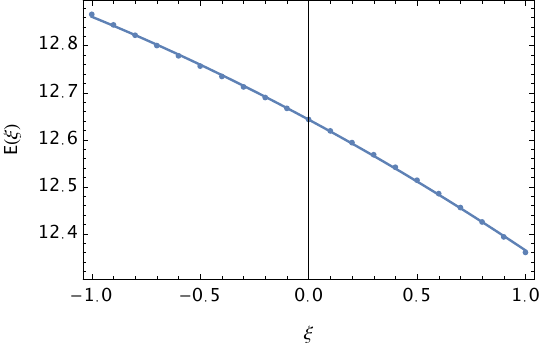} 
\caption{\label{dipole} The circle is the energy by traditional method for dipole interaction with $N=4, \beta=1$, and $\lambda=1$. The solid line is the parabolic function for fitting. By adding more parameters with $a_3 \xi^3$ and $a_4 \xi^4$, we do not notice perceptible difference. Hence, it is not necessary to add those extra curves in this figure.}
\end{center}
\end{figure}

In Fig. \ref{dipole}, we consider the  case of dipole interaction 
\begin{equation}
V_{int}=\sum_{l<j}^N \frac{\lambda}{{|{\textbf r}_l-{\textbf r}_j|^3}}
\end{equation} 
with $\lambda=1$ and $N=4,\beta=1$. For the calculated energy, we use the parabolic function for fitting and agreement is found with $\Delta=0.0024$. This shows again that $E(\beta,\xi)$ is a monotonic analytical function about $\xi$ for this situation.

\begin{figure}[htbp]
\begin{center}
 \includegraphics[width=0.9\textwidth]{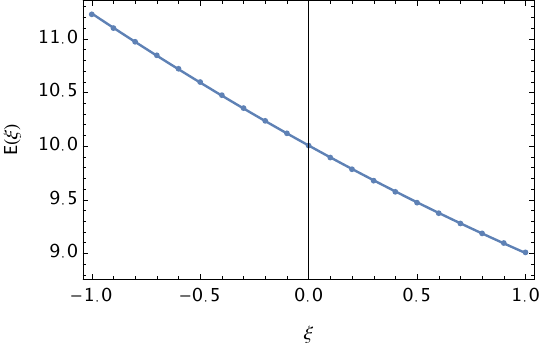} 
\caption{\label{Gauss} The circle is the energy by traditional method for Gaussian interaction with parameters  $g=3.0$, $s=0.5$, $N=4$, and $\beta=1$. The solid line is the parabolic function for fitting. By adding more parameters with $a_3 \xi^3$ and $a_4 \xi^4$, we do not notice perceptible difference. Hence, we do not add those extra curves in this figure.}
\end{center}
\end{figure}

In Fig. \ref{Gauss}, we consider the third example with Gaussian interaction 
\begin{equation}
V_{int}=\sum_{l<j}^N \frac{g}{\pi s^2}e^{-\frac{|{\textbf r}_l-{\textbf r}_j|^2}{s^2}}.
\end{equation}
The parameters are $g=3.0$, $s=0.5$ and $N=4,\beta=1$. The solid line gives the fitting with parabolic function, and agreement is found once again with $\Delta=0.0035$.

These examples show clearly that the average energy can have good analytical property for many different physical models. For $\xi\geq 0$, there is no sign problem. We have also checked that for all these examples, the addition of more parameters with $a_3 \xi^3$ and $a_4 \xi^4$ does not give perceptible difference in the fitting result.
This suggests strongly that we may calculate accurately the average energy with the computational scaling of $O(PN^3)$ for different values of positive $\xi$. Based on the assumption of the good analytical property of $E(\beta,\xi)$ in the whole region of $\xi\geq -1$, it is natural that we may get $E(\beta,\xi=-1)$ for fermions by a fitting of the data for $E(\beta,\xi\geq 0)$.

\section{Circumventing fermion sign problem by extrapolation}

The above study tells us that $U_\xi^{(N)}$ is a real function for $\xi\geq 0$, while a complex function for $\xi<0$. Therefore, in the whole real number region $\xi\geq 0$, the partition function for this fictitious quantum system does not suffer from sign problem. 
Assuming the analytical characteristic of $Z(\beta,\xi)$ and $E(\beta,\xi)$ about $\xi$, it is expected that the extrapolation will give us good chance to deal with fermions with $\xi=-1$. 
The accurate calculation of $E(\beta,\xi)$ for $\xi\geq 0$ means that we can obtain accurate evaluation of the analytical function $E(\beta,\xi)$. In this case, it's a matter of course that we have good chance to get accurate result of $E(\beta,\xi=-1)$ for fermions. 

We emphasize that the extrapolation to predict an unknown value from the fitting with known data is significantly different from the fitting for the known data demonstrated in Sec. \ref{fullbehavior}. Usually, in the fitting of data to predict the unknown value by extrapolation, 
the physical analysis and the experience and simplicity considerations play important role in the choice of the fitting function.
In Refs. \cite{HirshbergFermi,DornheimMod}, very simple functions are used for the fitting and extrapolation to attenuate the fermion sign problem. In practical research, another issue we should emphasize is that 
we should avoid both underfitting and overfitting \cite{Hawkins}. If the number of parameters has already given satisfying fitting, it is not necessary to add the parameters to improve the fitting, which may amplify the noice in the data for fitting. In particular, if the addition of extra parameter leads to obvious deviation to the prediction with less parameters, it is a clear signature of overfitting. Another signature of the overfitting is that the addition of extra parameters induces unexpected concave property or convex property to the fitting function. If the parameters chosen for fitting can not give good agreement with the data, it is an obvious underfitting, and more parameters should be used to improve the fitting and extrapolation. We will consider the situation of underfitting and overfitting by several examples, by the comparison with the accurate result in the traditional method.

For the examples considered in this paper, we find that the parabolic function 
$E(\beta,\xi)=a_0+a_1\xi+a_2\xi^2$ can already give us satisfying result. In principle, it is not necessary to add more parameters for further improvement, which may lead to possible overfitting. For the following examples, it is easy to find that the linear function $E(\beta,\xi)=a_0+a_1\xi$ will lead to obvious underfitting.

Nevertheless, for our seemingly natural idea, we must carry out numerical experiments to put it to the ultimate test. 
The evaluation of the precision of $E(\beta,\xi=-1)$ can be shown through the degree of conformity of all the data for $\xi\geq 0$ in the fitting, and the comparison with other method.
Now we turn to consider numerical experiments to test our method. In our numerical simulations, the path integral ring polymers and molecular dynamics with separate Nos\'e-Hoover thermostat \cite{Nose1,Nose2,Hoover,Martyna,Jang,Schlessinger} are used to sample the partition function $Z(\beta,\xi)$ for $\xi\geq 0$, which is further used by data fitting to calculate the energy for fermions. 
The number of beads $P$ used decreases as temperature increases to ensure numerical stability and assure convergence. In all of the following we checked convergence with respect to the number of beads and MD steps performed.

We consider the following Hamiltonian in two dimensions:
\begin{equation}
\hat H=\frac{1}{2m}\sum_{l=1}^N\hat \textbf p_l^2+\frac{1}{2}m\omega^2\sum_{l=1}^N \textbf r^2_l+\sum_{l<j}^N \frac{\lambda}{|{\textbf r}_l-{\textbf r}_j|}.
\end{equation}

\begin{figure}
\begin{center}
 \includegraphics[width=0.9\textwidth]{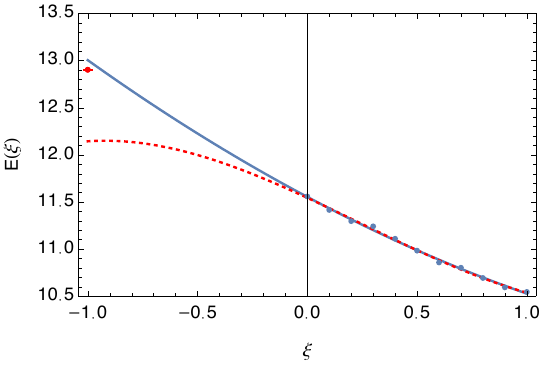} 
\caption{\label{4int} For $N=4, \beta=1$ and $\lambda=0.5$, the blue circle gives the energy for 
$\xi\geq 0$ based on PIMD. The solid line is the fitting with parabolic function. The red circle gives the result of traditional method in Ref. \cite{Dornheim}, which agrees with our result for fermions by fitting with parabolic function. If $a_3x^3$ is added as shown by the red dotted line, the convex property of $E(\xi)$ in the whole region is changed, which shows clearly an overfitting. The traditional method has already shown that $E(\xi)$ should be a convex function in the whole region.
Based on experience and simplicity, we think the fitting with parabolic function is appropriate in this example.
 }
\end{center}
\end{figure}

For $N=4, \beta=1$ and $\lambda=0.5$, the blue circle in Fig. \ref{4int} shows the numerical result of the energy for $0\leq \xi\leq 1$. The solid line is the best fitting with parabolic function.
The red circle is the result of path integral Monte Carlo in Ref. \cite{Dornheim}, which agrees with our result. The deviation with Dornheim's  result \cite{Dornheim} is $1\%$. For $\xi>0$, we also calculate the mean deviation of the fitting with the data, and find that $\Delta=0.014$.

\begin{figure}
\begin{center}
 \includegraphics[width=0.9\textwidth]{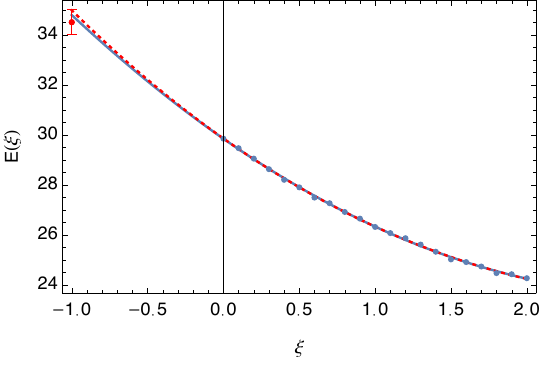} 
\caption{\label{8int} For $N=8, \beta=1$ and $\lambda=0.5$, the blue circle gives the energy for 
$\xi\geq 0$ based on PIMD. The solid line is the fitting with parabolic function. The red circle is the result of traditional method in Ref. \cite{Dornheim}. In the fitting with parabolic function, we get $E(\xi=-1)=34.79$, while the addition of $a_3 x^3$ gives $E(\xi=-1)=34.98$, which means the stability of our fitting in this example. The reason for this stability may be due to the fact that the inclusion of $a_3 x^3$ does not change the convex property of $E(\xi)$ in the whole region.
}
\end{center}
\end{figure}

In Fig. \ref{8int}, by doubling the particle number, we show our result for $N=8$, $\beta=1$ and $\lambda=0.5$, which agrees with the traditional method (red circle) by Dornheim \cite{Dornheim}, while our method is much more efficient and has smaller statistical fluctuations. Our result is well within the error bar of Ref.  \cite{Dornheim} with a deviation less than $1\%$. For $\xi>0$, the mean deviation of the fitting is $\Delta=0.04$.

\begin{figure}
\begin{center}
 \includegraphics[width=0.9\textwidth]{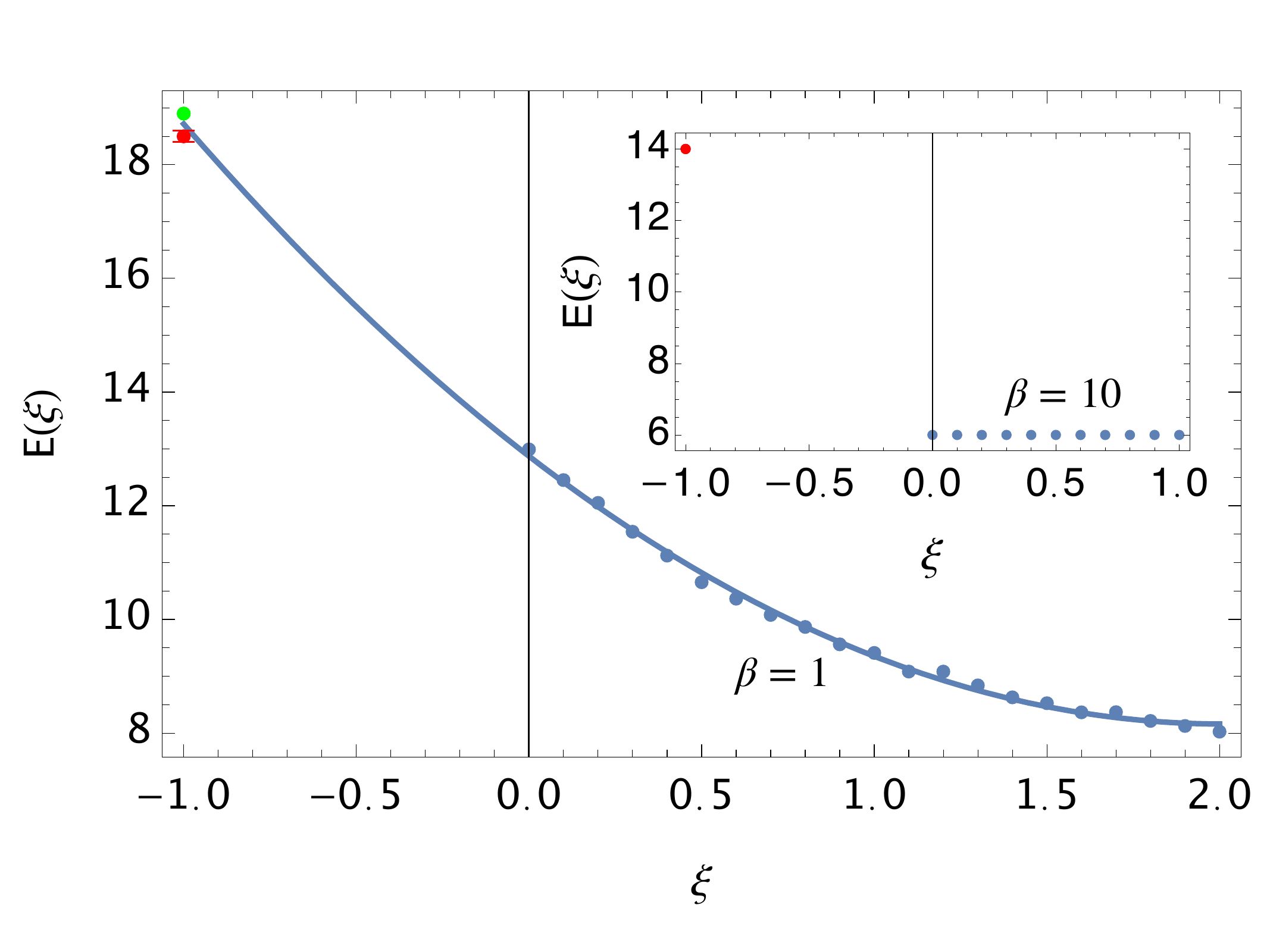} 
\caption{\label{6ideal} For noninteracting fermions with $N=6$ and $\beta=1$, the blue circle gives the result for $\xi\geq 0$, while the blue line gives the results by fitting with parabolic function. The red circle gives the result of traditional method \cite{Dornheim}, while the green circle gives the result of grand canonical ensemble. For this situation, the addition of $a_3 x^3$ will lead to obvious deviation, which means the overfitting for this situation. In this case, we believe the parabolic function is the appropriate choice in this example. The inset shows the example of noninteracting fermions with $N=6,\beta=10$. The red circle is the energy of fermions in the grand canonical ensemble, while the blue circle is the result of PIMD for different positive $\xi$. Almost constant energy for $\xi\geq 0$ shows clearly the invalidity of the extrapolation for this example.}
\end{center}
\end{figure}

We consider further an example of noninteracting fermions with $N=6$ and $\beta=1$, which is very difficult but feasible in traditional method. The Fermi temperature for this example is $T_F=3.5$ which is much larger than the temperature of $\beta=1$. It is worth pointing out that usually the noninteracting fermions at low temperatures below the Fermi temperature have severe fermion sign problem, which provide perfect system to check the method to overcome the fermion sign problem\cite{DornheimMod,HirshbergFermi}. 
In Fig. \ref{6ideal}, we give the numerical results for these parameters. The red circle shows the result of traditional method by path integral Monte Carlo \cite{Dornheim}, and the green circle gives the result of grand canonical ensemble. The deviation with Dornheim's result \cite{Dornheim} is $1\%$. For $\xi>0$, the mean deviation of the fitting is $\Delta=0.09$.

Although we have obtained good result for this example of noninteracting fermions, it is worth pointing out that our method can not be applied to zero temperature or extremely low temperature of noninteracting fermions so that $E(\xi)$ has no obvious change for different positive $\xi$. For the extreme situation of zero temperature, all particles are in the ground state for $0\leq \xi\leq 1$. In this case, the energy $E(\xi)$ is a constant for $0\leq \xi\leq 1$. It is impossible for us to get the correct energy by extrapolation for this situation. For $N=6,\beta=10$ as an example shown in the inset of Fig. \ref{6ideal}, our calculations show that we can not get the correct energy for this situation.

It is worth pointing out that for $\xi\geq 0$, by choosing the parameters for massive Nóse-Hoover chain appropriately the error can be made small and independent of temperature.
When the interparticle interaction is included, however, the different exchange interaction for different positive $\xi$ will make $E(\xi)$ depend on the parameter $\xi$, so that we may have the chance to get the correct energy for fermions by extrapolation. We will test the validity of this conjecture for zero-temperature interacting fermions in future research. 

In all the results based on our method, the statistical error is negligible. Based on molecular dynamics experiences $5\times 10^6$ MD steps is enough to obtain accurate results for most systems. Hence, we do not give error bar for the energy in our method. We have checked for some examples (including our previous works of path integral molecular dynamics for bosons \cite{Xiong,Xiong4}) that for ten independent trajectories of much more MD steps, we do not notice perceptible difference. This is different from the calculation of the energy for fermions by traditional method. Of course, this difference is due to the difficulty in traditional method for fermions.
The negligible statistical error for the calculation of the energy for $\xi\geq 0$  provides strict criteria to test whether our method is applicable for specific and eccentric quantum systems in future work. From all these examples, it seems that the deviation $\Delta E(\xi=-1)$ is one order of magnitude greater than $\Delta$ for $
\xi>0$. Of course, this is the price to pay for extrapolation. Fortunately, for $\xi>0$ without suffering from fermion sign problem it is not difficult to get the precision of $0.1\%$, which means that we have good chance of obtaining the energy for fermions with the precision of $1\%$, which is good enough for many practical applications.

The successful application of our method to fermion systems is not so surprising.
Because the operator $e^{-\beta\hat H}$ is hermitian, it is clear that $Z(\beta,\xi)$ is a real function of $\beta$ and $\xi$ even for $\xi=-1$. For fermions, the transformation of $Z(\beta,\xi)$ as a multiple integral expression by path integral ring polymer is an important merit in the traditional method. However, after this transformation, for $\xi=-1$ (fermions), $U_F^{(N)}$ in the exponential function $e^{-\beta U_F^{(N)}}$ in the partition function becomes a complex function. This is a huge cost to study fermions by path integral ring polymer. However, we should never forget that $Z(\beta,\xi)$ in essence is a problem of real number, and the emergence of complex exponential function and the resulting fermion sign problem is due to specific mathematical technique.

It is not the purpose of the present work to prove or argue that our polynomial-time algorithm can be used to study any fermion system. The present method relies on the assumption that $E(\beta,\xi)$ has analytical property for $\xi\geq -1$. 
 Of course, for some specific quantum systems, we may find better function than polynomial function for data fitting. It depends on the quantum system studied, and the  comparison with experiments may also improve the choice of the fitting function.

\section{Summary and discussion}

\begin{figure}
\begin{center}
 \includegraphics[width=0.9\textwidth]{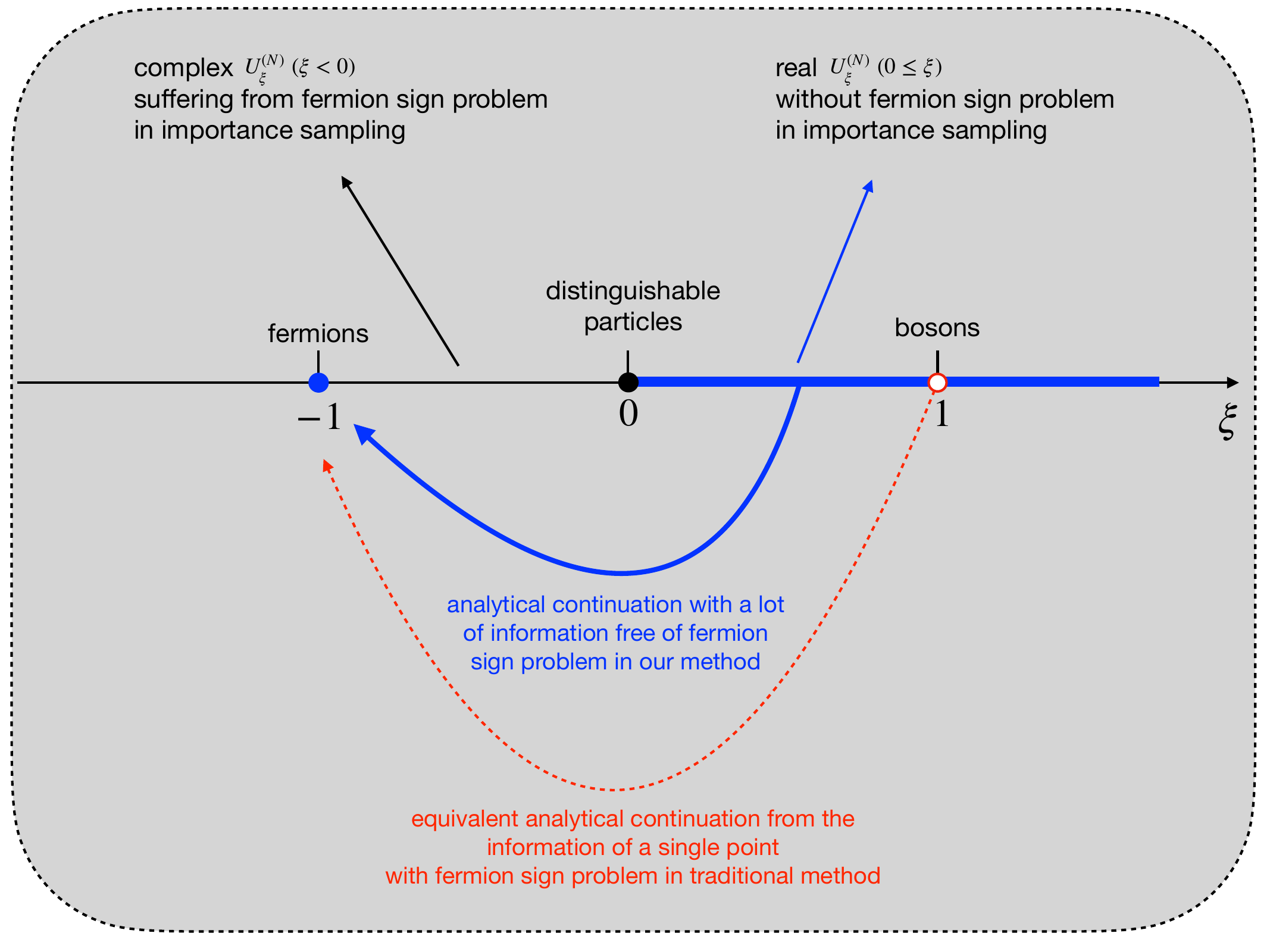} 
\caption{\label{AnalyticalC} For the partition function $Z(\beta,\xi)$ with $\xi$ a real number ($\xi=-1$ for fermions, $\xi=0$ for distinguishable particles and $\xi=1$ for bosons), $U_\xi^{(N)}$ in the whole region $\xi\geq 0$ is a real function, which can be sampled accurately and efficiently. Hence, we can get many reliable values for the energy $E(\beta,\xi)$ in this region, and analytically continue to the energy of fermions ( illustrated by blue solid line with arrow) if $E(\beta,\xi)$ has simple analytical property about $\xi$. In the traditional method, however, it is equivalent that the distribution of the partition function for a single point $\xi=1$ is used to try to extend to the energy for fermions illustrated by red dashed line, which leads to the computational scaling of $O(e^{\beta N})$.}
\end{center}
\end{figure}

Compared with the traditional method, it is shown in this work that we have the chance to circumvent the finite-temperature fermion sign problem with much higher efficiency. 
The significant increasing of the efficiency is due to the fact that in our method, more information (the reliable importance sampling for many different values of $\xi\geq 0$ for $Z(\beta,\xi)$) are given to the analytical property of $Z(\beta,\xi)$ to extend the result to $\xi=-1$ (See Fig. \ref{AnalyticalC}). As a comparison, we see that in the traditional method,  we only consider the distribution of the partition function $Z(\beta,\xi)$ for a single point $\xi=1$ and try to analytically extend to the result to $\xi=-1$.

Because in our method, the energy for $\xi\geq 0$ can be calculated accurately (it is routine to realize the precision far below $1\%$), the only potential problem to get the accurate value for fermions is whether the fitting is reasonable. The good fitting depends on the good analytical property of $E(\beta,\xi)$ between $-1\leq\xi\leq 1$.  
In this work, we use Mathematica for fitting, while the energy between $0\leq\xi\leq 1$ is calculated by path integral molecular dynamics with C++ codes. In a sense, the choice of polynomial function for fitting in this paper is based on empirical simplicity consideration, and the fact that Taylor series is a good approximation for many analytical functions. We can not exclude the possibility in some quantum systems, the polynomial function probably wouldn't be a good choice.
If $E(\beta,\xi)$ has singular behavior or highly irregular oscillating behavior about $\xi$, it is hard to assure the extrapolation for fermions by polynomial function. However, even for these situations, some unique methods such as Schlessinger point method \cite{Schlessinger} may be helpful. Of course, for any practical research, we should first check the numerical stability and accuracy for $\xi\geq 0$ and judge the validity of the fitting also from physical considerations. Before considering larger fermion systems, one may first calculate a few fermions for the region $-1\leq\xi\leq 1$ by traditional method to confirm the simple and monotonic behavior of $E(\beta,\xi)$. After this confirmation, it would be safe to get the energy for larger fermion systems using the data fitting for $\xi\geq 0$. It is expected that the present method would be an alternative to  restricted path integral Monte Carlo \cite{nodes,Helium,Militzer}, density matrix quantum Monte Carlo \cite{Blunt,Malone}, configuration path integral Monte Carlo \cite{Schoof1, Schoof2,Schoof3, Yilmaz}, and auxiliary field quantum Monte Carlo \cite{Joonho} for fermion sign problem at finite temperatures.

\begin{acknowledgments}
This work is partly supported by the National Natural Science Foundation of China under grant numbers 11175246, and 11334001. We acknowledge useful suggestion to improve the paper by X. S. Chen, 
X. Z. Wang and X. Q. Lin. We acknowledge in particular the encouragement and important comment to improve the paper by Frank Wilczek.
\end{acknowledgments}

\textbf{DATA AVAILABILITY}

The data that support the findings of this study are available from the corresponding author upon reasonable request. The code of this study is openly available in GitHub (https://github.com/xiongyunuo/PIMD-Pro).


\begin{thebibliography}{10}


\bibitem{feynman} R. P.~Feynman and A. R.~Hibbs, Quantum mechanics and path integrals, Dover Publications, New York (2010).

\bibitem{kleinert} H.~Kleinert, Path integrals in quantum mechanics, statistics, polymer physics, and financial markets, World Scientific, Singapore (2009).

\bibitem{Tuckerman} M. E.~Tuckerman, Statistical mechanics: theory and molecular simulation, Oxford University, New York (2010).


 \bibitem{chandler} D.~Chandler and P. G.~Wolynes, Exploiting the isomorphism between quantum theory and classical statistical mechanics of polyatomic fluids, {\text{J.~Chem.~Phys.}~\textbf{74}, 4078} (1981).

\bibitem{Parrinello} M.~Parrinello and A.~Rahman, Study of an F center in molten KCl, \text{J. Chem. Phys.}~\textbf{80}, 860 (1984).

\bibitem{Miura} S. Miura and S. Okazaki, Path integral molecular dynamics for Bose-Einstein and Fermi-Dirac statistics. \text{J. Chem. Phys.} \textbf{112}, 10116 (2000).

 \bibitem{Cao} J. Cao and G. A. Voth,  The formulation of quantum statistical mechanics based on the Feynman path centroid density. I. Equilibrium properties, \text{J. Chem. Phys.} \textbf{100},  5093 (1994).
 
 \bibitem{Cao2} J. Cao and G. A. Voth, The formulation of quantum statistical mechanics based on the Feynman path centroid density. II. Dynamical properties,  \text{J. Chem. Phys.} \textbf{100}, 5106 (1994).
 
 \bibitem{Jang2} S. Jang and G. A. Voth, A derivation of centroid molecular dynamics and other approximate time evolution methods for path integral centroid variables,  \text{J. Chem. Phys.} \textbf{111}, 2371 (1999).
 
\bibitem{Ram} R. Ram\'iRez and T. L\'oPez-Ciudad, The Schr\"odinger formulation of the Feynman path centroid density, \text{J. Chem. Phys.} \textbf{111}, 3339 (1999).


\bibitem{Kinugawa} K. Kinugawa, H. Nagao, and K. Ohta, Path integral centroid molecular dynamics method for Bose and Fermi statistics: formalism and simulation, Chem. Phys. Lett. \textbf{307}, 187 (1999).

\bibitem{Roy} Pierre-Nicholas Roy, Seogjoo Jang, and Gregory A. Voth, Feynman path centroid dynamics for Fermi–Dirac statistics, J. Chem. Phys. \textbf{111}, 5303 (1999).

\bibitem{Roy1} P.-N. Roy and G.A. Voth, On the Feynman path centroid density for Bose-Einstein and Fermi-Dirac statistics, J. Chem. Phys. \textbf{110}, 3647 (1999).


\bibitem{Roy2} N. Blinov, P.-N. Roy, and G.A. Voth, Path integral formulation of centroid dynamics for systems obeying Bose-Einstein statistics, J. Chem. Phys. \textbf{115}, 4484 (2001).

\bibitem{Roy3} N. Blinov and P.-N. Roy, Operator formulation of centroid dynamics for Bose-Einstein and Fermi-Dirac statistics, J. Chem. Phys. \textbf{115}, 7822 (2001).

\bibitem{Kinugawa1} K. Kinugawa, A semiclassical approach to the dynamics of many-body Bose/Fermi systems by the path integral centroid molecular dynamics, J. Chem. Phys. \textbf{114}, 1454 (2001).

\bibitem{Kinugawa2} K. Kinugawa, H. Nagao, and K. Ohta, A path integral centroid molecular dynamics method for Bose and Fermi statistics, J. Mol. Liq. \textbf{90}, 11 (2001).

\bibitem{Roy4} Nicholas Blinov and Pierre-Nicholas Roy, An effective centroid Hamiltonian and its associated centroid dynamics for indistinguishable particles in a harmonic trap, J. Chem. Phys. \textbf{116}, 4808 (2002).

\bibitem{Roy5} Pierre-Nicholas Roy and Nicholas Blinov, Centroid dynamics with quantum statistics, Isr J. Chem. \textbf{42}, 183 (2002).

\bibitem{Roy6} Paul Moffatt, Nicholas Blinov, and 
Pierre-Nicholas Roy, On the calculation of single-particle time correlation functions from Bose–Einstein centroid dynamics, J. Chem. Phys. \textbf{120}, 4614 (2004).

 
\bibitem{Poly} E. A. Polyakov, A. P.  Lyubartsev, and P. N.  Vorontsov-Velyaminov,  Centroid molecular dynamics: Comparison with exact results for model systems, \text{J. Chem. Phys.} \textbf{133}, 194103 (2010). 

 \bibitem{Craig} I. R. Craig and D. E. Manolopoulos, Quantum statistics and classical mechanics: Real time correlation functions from ring polymer molecular dynamics,  \text{J. Chem. Phys.}  \textbf{121}, 3368 (2004). 
 
 \bibitem{Braa} B. J. Braams and D. E. Manolopoulos, On the short-time limit of ring polymer molecular dynamics, \text{J. Chem. Phys.} \textbf{125}, 124105 (2006).
 
 \bibitem{Haber} S. Habershon, D. E. Manolopoulos, T. E. Markland, and T. F. Miller 3rd, Ring-polymer molecular dynamics: quantum effects in chemical dynamics from classical trajectories in an extended phase space, \text{Annu. Rev. Phys. Chem.} \textbf{64}, 387 (2013).
 
 \bibitem{Thomas} T. E. Markland and M. Ceriotti, Nuclear quantum effects enter the mainstream, \text{Nat. Rev. Chem.} \textbf{2}, 0109 (2018).
 
\bibitem{CeperRMP} D. M. Ceperley, 
Path integrals in the theory of condensed helium, 
\text{Rev. Mod. Phys.} \textbf{67}, 279 (1995).

\bibitem{boninsegni1} M.~Boninsegni, N. V.~Prokof’ev, and B. V.~Svistunov, 
Worm algorithm and diagrammatic Monte Carlo: A new approach to continuous-space path integral Monte Carlo simulations, 
 {\text{Phys.~Rev.~E}~\textbf{74}, 036701} (2006).

\bibitem{boninsegni2} M.~Boninsegni, N. V.~Prokof’ev, and B. V.~Svistunov, 
Worm algorithm for continuous-space path integral Monte Carlo simulations,  
{\text{Phys.~Rev.~Lett.}~\textbf{96}, 070601} (2006).

\bibitem{Dornheim} T.~Dornheim, 
The Fermion sign problem in path integral Monte Carlo simulations: quantum dots, ultracold atoms, and warm dense matter, 
\text{Phys. Rev. E}~\textbf{100}, 023307 (2019).

\bibitem{DornheimMod} T. Dornheim, M. Invernizzi, J. Vorberger, and B. Hirshber, 
Attenuating the fermion sign problem in path integral Monte Carlo simulations using the Bogoliubov inequality and thermodynamic integration, 
J. Chem. Phys. \textbf{153}, 234104 (2020).



\bibitem{Hirshberg} B. Hirshberg, V. Rizzi, and M. Parrinello, Path integral molecular dynamics for bosons, 
\text{Proc. Natl. Acad. Sci. U. S. A.}~\textbf{116}, 21445 (2019).

\bibitem{HirshbergFermi} B. Hirshberg,  M. Invernizzi, and  M. Parrinello, 
Path integral molecular dynamics for fermions: Alleviating the sign problem with the Bogoliubov inequality, 
\text{J. Chem. Phys.} \textbf{152}, 171102 (2020).

\bibitem{Deuterium}   C. W. Myung, B. Hirshberg, and M. Parrinello, 
Prediction of a supersolid phase in high-pressure deuterium, 
\text{Phys. Rev. Lett.} \textbf{128}, 045301 (2022).

\bibitem{Xiong} Y. N. Xiong and H. W. Xiong, 
Path integral molecular dynamics simulations for Green's function in a system of identical bosons, 
J. Chem. Phys. \textbf{156}, 134112 (2022).

\bibitem{Xiong2} Y. N. Xiong and  H. W. Xiong, 
Numerical calculation of Green's function and momentum distribution for spin-polarized fermions by path integral molecular dynamics, 
J. Chem. Phys. \textbf{156}, 204117 (2022).

\bibitem{Xiong3} Y. N. Xiong and  H. W. Xiong, Path integral and winding number in singular magnetic field, Eur. Phys. J. Plus \textbf{137}, 550 (2022).

\bibitem{Xiong4} Y. L. Yu, S. J. Liu, H. W. Xiong, and Y. N. Xiong, Path integral molecular dynamics for thermodynamics and Green's function of ultracold spinor bosons, arXiv:2207.07653 (2022).

\bibitem{Xiong5} Y. N. Xiong and H. W. Xiong, Path integral molecular dynamics for anyons, bosons and fermions, arXiv:2207.11252 (2022).

\bibitem{ceperley} D. M.~Ceperley, Path Integral Monte Carlo Methods for Fermions, Monte Carlo and Molecular Dynamics of Condensed Matter Systems, K.~Binder and G.~Ciccotti (Eds.), Bologna (Italy) (1996).

\bibitem{Alex} A. Alexandru, G. Basar, P. F. Bedaque, and N. C. Warrington, 
Complex paths around the sign problem, 
Rev. Mod. Phys. \textbf{94}, 015006 (2022).

\bibitem{troyer} M.~Troyer and U. J.~Wiese, 
Computational Complexity and Fundamental Limitations to Fermionic Quantum Monte Carlo Simulations, 
{\text{Phys. Rev. Lett.} \textbf{94}, 170201} (2005).


\bibitem{loh} E. Y.~Loh, J. E.~Gubernatis, R. T.~Scalettar, S. R.~White, D. J.~Scalapino, and R. L.~Sugar, 
Sign problem in the numerical simulation of many-electron systems, 
{\text{Phys. Rev. B} \textbf{41}, 9301} (1990).


\bibitem{lyubartsev} A. P.~Lyubartsev, 
Simulation of excited states and the sign problem in the path integral Monte Carlo method, 
{\text{J.~Phys.~A: Math.~Gen.}~\textbf{38}, 6659} (2005).

\bibitem{vozn} M. A.~Voznesenskiy, P. N.~Vorontsov-Velyaminov, and A. P.~Lyubartsev, 
Path-integral-expanded-ensemble Monte Carlo method in treatment of the sign problem for fermions, 
\text{Phys.~Rev.~E}~\textbf{80}, 066702 (2009).


\bibitem{Science} R. Mondaini, S. Tarat, and R. T. Scalettar,  Quantum critical points and the sign problem, Science \textbf{375}, 
418 (2022).


\bibitem{Wu} Congjun Wu and Shou-Cheng Zhang, 
Sufficient condition for absence of the sign problem in the fermionic quantum Monte Carlo algorithm, 
Phys. Rev. B \textbf{71}, 155115 (2005).

\bibitem{Umrigar} C. J. Umrigar, Julien Toulouse, Claudia Filippi, S. Sorella, and R. G. Hennig, 
Alleviation of the Fermion-Sign Problem by Optimization of Many-Body Wave Functions, 
Phys. Rev. Lett. \textbf{98}, 110201 (2007).

\bibitem{Li} Zi-Xiang Li, Yi-Fan Jiang, and Hong Yao, 
Solving the fermion sign problem in quantum Monte Carlo simulations by Majorana representation, 
Phys. Rev. B \textbf{91}, 241117(R) (2015).

\bibitem{Wei} Z. C. Wei, Congjun Wu, Yi Li, Shiwei Zhang, and T. Xiang, 
Majorana Positivity and the Fermion Sign Problem of Quantum Monte Carlo Simulations, 
Phys. Rev. Lett. \textbf{116}, 250601 (2016).




\bibitem{Yao1} Z. X.~Li, Y.~F. Jiang, and H.~Yao, Solving the fermion sign problem in quantum Monte Carlo simulations by Majorana representation, {\text{Phys.~Rev.~B} \textbf{91}, 241117(R)} (2015).

\bibitem{Yao2} Z. X.~Li, Y.~F. Jiang, and H.~Yao, Majorana-Time-Reversal Symmetries: A Fundamental Principle for Sign-Problem-Free Quantum Monte Carlo Simulations, { \text{Phys.~Rev.~Lett.}~\textbf{117}, 267002} (2016).

\bibitem{Griffiths} David J. Griffiths and  Darrell F. Schroeter, Introduction to Quantum Mechanics, Pearson Education, Cambridge (2018).


\bibitem{Nose1} S. Nos\'e, 
A molecular dynamics method for simulations in the canonical ensemble, 
\text{Mol. Phys.} \textbf{52}, 255 (1984).

\bibitem{Nose2} S. Nos\'e, 
A unified formulation of the constant temperature molecular dynamics methods, 
\text{J. Chem. Phys.} \textbf{81}, 511 (1984).

\bibitem{Hoover} W. G. Hoover, 
Canonical dynamics: Equilibrium phase-space distributions, 
\text{Phys. Rev. A} \textbf{31}, 1695 (1985).

\bibitem{Martyna} G. J. Martyna, M. L. Klein, and M. Tuckerman, 
Nos\'e-Hoover chains: The canonical ensemble via continuous dynamics, 
\text{J. Chem. Phys.} \textbf{97}, 2635 (1992).

\bibitem{Jang} S. Jang and G. A. Voth, 
Simple reversible molecular dynamics algorithms for Nos\'e-Hoover chain dynamics, 
\text{J. Chem. Phys.}~\textbf{107}, 9514 (1997).

\bibitem{Schlessinger} L. Schlessinger, Use of Analyticity in the Calculation of Nonrelativistic Scattering Amplitudes, Phys. Rev. \textbf{167}, 1411 (1968).

\bibitem{Hawkins} Douglas M. Hawkins, The Problem of Overfitting, J. Chem. Inf. Comput. Sci.  \textbf{44}, 1 (2004).




\bibitem{nodes} D. M. Ceperley, Fermion nodes, J. Stat. Phys., \textbf{63}, 1237 (1991).

\bibitem{Helium}  D. M. Ceperley, Path-integral calculations of normal liquid 
3He, Phys. Rev. Lett. \textbf{69}, 331 (1992).

\bibitem{Militzer} B. Militzer, E. L. Pollock, and D. M. Ceperley, Path integral Monte Carlo calculation of the momentum distribution of the homogeneous electron gas at finite temperature, High Energy Dens. Phys. \textbf{30}, 13 (1029).


\bibitem{Blunt} N. S. Blunt, T. W. Rogers, J. S.  Spencer, and W. M. Foulkes,
Density-matrix quantum Monte Carlo method, Phys.
Rev. B \textbf{89}, 245124 (2014). 

\bibitem{Malone} F.D. Malone, N. S. Blunt, James J. Shepherd, D. K. K. Lee,  J. S. Spencer, and  W. M. C. Foulkes, Interaction Picture Density Matrix Quantum Monte Carlo, J. Chem. Phys. \textbf{143}, 044116 (2015).


\bibitem{Schoof1} T. Schoof, M. Bonitz, A. V. Filinov, D. Hochstuhl and
J. W. Dufty, Configuration Path Integral Monte Carlo,
Contrib. Plasma Phys. \textbf{51}, 687 (2011).

\bibitem{Schoof2} T. Schoof, S. Groth,  and M. Bonitz, Towards ab Initio Thermodynamics of the Electron Gas at Strong Degeneracy, Contrib. Plasma Phys. \textbf{55}, 136 (2015).

\bibitem{Schoof3} T. Schoof, S. Groth, J. Vorberger,  and M. Bonitz, Ab Initio Thermodynamic Results for the Degenerate Electron Gas at Finite Temperature, Phys. Rev. Lett. \textbf{115}, 130402 (2015).

\bibitem{Yilmaz} A. Yilmaz, K. Hunger,  T. Dornheim, S. Groth, and  M. Bonitz, Restricted configuration path integral Monte Carlo, J. Chem. Phys. \textbf{153}, 124114 (2020).


\bibitem{Joonho} Joonho Lee, Miguel A. Morales, and Fionn D. Malone, A phaseless auxiliary-field quantum Monte Carlo perspective on the uniform electron gas at finite temperatures: Issues, observations, and benchmark study, J. Chem. Phys. \textbf{154}, 064109 (2021).


\end{thebibliography}
\end{document}